\documentclass[english,prb, twocolumn]{revtex4}
\usepackage[T1]{fontenc}
\usepackage[latin9]{inputenc}
\usepackage{amsmath}
\usepackage{graphicx}
\usepackage{amssymb}

\makeatletter
\@ifundefined{textcolor}{}
{%
 \definecolor{BLACK}{gray}{0}
 \definecolor{WHITE}{gray}{1}
 \definecolor{RED}{rgb}{1,0,0}
 \definecolor{GREEN}{rgb}{0,1,0}
 \definecolor{BLUE}{rgb}{0,0,1}
 \definecolor{CYAN}{cmyk}{1,0,0,0}
 \definecolor{MAGENTA}{cmyk}{0,1,0,0}
 \definecolor{YELLOW}{cmyk}{0,0,1,0}
 }

\usepackage{color}

\makeatother

\usepackage{babel}

\begin{document}

\title{Polaron and bipolaron transport in a charge segregated state of doped
strongly correlated 2D semiconductor}

\author{J. Miranda, T. Mertelj, V. Kabanov, D. Mihailovic}

\affiliation{Complex Matter Dept., Jozef Stefan Institute, Jamova 39, Ljubljana,
SI-1000, Ljubljana, Slovenia }
\begin{abstract}
The 2D lattice gas model with competing short and long range interactions
is used for calculation of the incoherent charge transport in the
classical strongly-correlated charge segregated polaronic state. We
show, by means of Monte-Carlo simulations, that at high temperature
the transport is dominated by hopping of the dissociated correlated
polarons, where their mobility is inversely proportional to the temperature,
$\mu\propto T^{-1}$. At temperatures below the clustering transition
temperature the bipolaron transport becomes dominant. The energy barrier
for the bipolaron hopping is determined by Coulomb effects and is
found to be lower than the barrier for the single-polaron hopping.
This leads to drastically different temperature dependencies of mobilities
for polarons and bipolarons at low temperatures. 
\end{abstract}

\date{\today}

\maketitle

\section{Introduction}

The physics of small and large polarons is increasingly one of the
most intriguing branches of contemporary solid state physics. Polarons
are often discussed in connection to the transport and optical properties
of doped semiconductors\cite{alexandrovdevrees,polinadvaced}. The
kinetic properties of polaronic systems are well studied in the limit
of small density of polarons\cite{Firsov}. However, the situation
becomes more complicated in the case of heavily doped semiconductors
and insulators. The correlations between polarons may lead to the
formation of a homogeneous gas of bipolarons or the formation of a
charge ordered\cite{alexandrov1986} or phase separated state\cite{mertelj}.
Indeed charge segregation of polarons has recently been discussed
in literature\cite{manganites}. The kinetic properties of a dilute
bipolaron gas are known in the band limit\cite{alemott} and in the
limit of hopping conductivity\cite{bryksin}. On the other hand the
kinetic properties of charge ordered and especially phase separated
polarons and bipolarons remain an open and challenging problem.

The generally accepted mechanism of charge transport in doped semiconductors
at low temperature is variable range hopping (VRH) of localized electrons
in the impurity band\cite{mottdavis}. Inclusion of the long range
Coulomb interaction leads to the opening of the Efros-Shkolvskii Coulomb
gap and a modified temperature dependence of the conductivity, compared
to the VRH model\cite{efros}. Increasing the complexity further,
there have been increasing debates about the role of multi-electron
hopping conductivity. Tielsen and Schreiber have performed extensive
calculations and came to conclusion that the many-electron hopping
processes become dominant at low temperatures\cite{tenelsen}, which
was also supported by simulations\cite{pollak1}. Later, these results
were challenged by Tsigankov and Efros\cite{tsigankov} who argued
that a different choice of hopping rates leads to a substantial suppression
of the two-particle contribution. More recently\cite{pollak2} it
was shown that if one assumes more general hopping rates, the many-particle
contribution to the conductivity may remain important at low temperatures.
Thus the many-particle contribution to the conductivity of doped semiconductors
and insulators, particularly at low temperatures remains controversial
and limits our understanding of the transport not only in doped polar
semiconductors, but also doped cuprates (including high-temperature
superconductors)\cite{hightc}, doped manganites\cite{manganites}
and many other materials\cite{elastic} where the electron-phonon
interaction is intermediate or strong.

Recently we have shown that doping a semiconductor with charged carriers
leads to the segregation of polarons on short length scale\cite{mertelj}.
We have shown that polaron clustering with an even number of particles
is more favorable than with an odd number of particles, because of
the particular symmetry of the short range potential and topology
of the lattice. This suggests that we can consider these clusters
to be effectively composed from bipolarons and it is natural to expect
that the low temperature conductivity will be determined by bipolaron
hopping because the formation of clusters with an odd number of polarons
are energetically less favorable. In our model calculations\cite{mertelj}
the distance between polarons within the bipolaron is small (inter-site
bipolarons) so the model calculations of electronic transport are
less sensitive to the choice of the hopping probability\cite{tsigankov}
than in the case of the standard Coulomb gas.

Here we present extensive numerical simulations of polaron\cite{MirandaMertelj2009}
and bipolaron hopping conductivity in the charge segregated state
which appears as a result of doping the strongly-correlated insulator.
We show that at low concentration and low temperature bipolaron transport
becomes dominant whereby the addition of a short range attraction
caused by the lattice deformation to the standard Coulomb gas Hamiltonian
unambiguously leads to the dominant role of the two-particle conductivity
at low temperatures.

\section{The model}

To perform the simulations we chose the lattice gas model on a 2D
square lattice with competing long-range Coulomb and short range anisotropic
Jahn-Teller (JT)-like interactions, because it naturally describes
the formation of bipolarons and bipolaronic clusters. The model in
terms of the pseudospin operators, $S_{\mathbf{i}}^{z}$, can be written
as,\cite{mertelj,MerteljKabanov2007}

\begin{equation}
H=\sum_{\textbf{i,j}}(V_{\mathrm{JT}}\left(\mathbf{i-j}\right)S_{\textbf{i}}^{z}S_{\textbf{j}}^{z}+V_{\mathrm{C}}\left(\mathbf{i}-\mathbf{j}\right)Q_{\textbf{i}}Q_{\textbf{j}}),\label{eq:hamilt}\end{equation}
where $S_{\mathbf{i}}^{z}=1,$ $S_{\mathbf{i}}^{z}=-1$ represent
a single particle in either of the states of the electronic JT doublet
at the site $\mathbf{i}$ and $S_{\mathbf{i}}^{z}=0$ represents an
unoccupied site. $Q_{\textbf{i}}=(S_{\textbf{i}}^{z})^{2}$ is the
on-site charge and $V_{C}\left(\mathbf{m}\right)=1/m$ is the dimensionless
Coulomb potential. $V_{JT}\left(\mathbf{i-j}\right)$ represents the
anisotropic short range JT interaction\cite{mertelj,MerteljKabanov2007}.
This Hamiltonian was used to demonstrate the stability of a variety
of different textures which might appear upon doping of strongly correlated
2D insulators. It was shown\cite{mertelj,MerteljKabanov2007} that
the short range attractive potential alone leads to global phase separation
which is frustrated by the long-range Coulomb potential\cite{MerteljKabanov2007,jamei}.
The system has a glassy ground state with many minima in the free
energy.

\subsection{Single polaron hopping}

To analyze the charge transport we use standard variable range hopping
transport theory\cite{LevinNguen1987}. The applicability of this
theory is justified by the glassy ground state demonstrated by previous
simulations\cite{mertelj,MerteljKabanov2007}. The \emph{single particle
hopping rate} between an occupied site $\mathbf{i}$ with pseudospin
$\alpha$ to an empty site $\mathbf{j}$ with the resulting pseudospin
$\beta$ is given by:\cite{LevinNguen1987} \begin{equation}
\gamma_{\mathbf{i}\alpha,\mathbf{j}\beta}^{\mathrm{sp}}=\gamma_{0}e^{-2\alpha_{\mathrm{sp}}r_{\mathbf{ij}}}\begin{cases}
e^{-\frac{\Delta_{\mathbf{i}\alpha,\mathbf{j}\beta}}{T}} & \mathrm{;\,}\Delta_{\mathbf{\mathbf{i}\textnormal{\ensuremath{\alpha},}j}\beta}>0\\
1 & \mathrm{;}\,\Delta_{\mathbf{\mathbf{i\textnormal{\ensuremath{\alpha},}j\textnormal{\ensuremath{\beta}}}}}\leq0\end{cases},\label{eq:hop-prob}\end{equation}
where $\Delta_{\mathbf{i}\alpha,\mathbf{j}\beta}=\Delta H\left(\mathbf{i\textnormal{\ensuremath{\alpha\rightarrow}}j\textnormal{\ensuremath{\beta}}}\right)-Ex_{\mathbf{ij}}$
is the dimensionless many body energy difference between the final
and initial states in the presence of an external electric field,
$E$, applied along the $x$-axis. $T$ is the temperature and $\gamma_{0}$$\exp\left(-2\alpha_{\mathrm{sp}}r_{\mathbf{ij}}\right)$
is the transition probability, $r_{\mathbf{ij}}=|\mathbf{j}-\textbf{i}|$
and $\alpha_{\mathrm{sp}}^{-1}$ the spatial extent of the single
polaron wavefunction. For a given system configuration, $\left\{ S_{\mathbf{m}}^{z}\right\} $,
the total dimensionless instantaneous current is the sum over all
possible hops: \begin{equation}
i^{\mathrm{sp}}=\sum\limits _{\textbf{i}\alpha}\sum\limits _{\textbf{j\ensuremath{\beta}}}x_{\mathbf{ij}}\gamma_{\mathbf{i\alpha,j\beta}}^{\mathrm{sp}}\,,\end{equation}
where $x_{\mathbf{ij}}=j_{x}-i_{x}$, with the first sum running over
all occupied states with $S_{\mathbf{i}}^{z}\ne0$ and the second
over all empty sites with $S_{\mathbf{j}}^{z}=0$ with both possible
final pseudospin states, $\beta=\pm1$. Since only the hops with $\Delta_{\mathbf{\mathbf{i\alpha}},\mathbf{\mathbf{j\beta}}}>0$
contribute to the net current the formula simplifies in the limit
of the vanishing electric field to: \begin{equation}
i^{\mathrm{sp}}=N_{\mathrm{eff}}\mu_{\mathrm{sp}}E\approx\gamma_{0}\frac{E}{T}\sum\limits _{\textbf{i}\alpha}\sum\limits _{\textbf{j\ensuremath{\beta}}}^{\prime}x_{\mathbf{ij}}^{2}e^{-2\alpha_{\mathrm{sp}}r_{\mathbf{ij}}}e^{-\frac{\Delta H\left(\mathbf{i\alpha\rightarrow j\beta}\right)}{T}},\label{eq:sp-current}\end{equation}
where $N_{\mathrm{eff}}$ is the total number of particles/holes%
\footnote{For fillings where more than half of the sites are occupied we use
the hole-polaron mobility for convenience so $N_{\mathrm{eff}}$ represents
the number of holes in the system.%
} and $\mu_{\mathrm{sp}}$ the dimensionless single polaron mobility.
The second sum runs only over the final states with $\Delta H\left(\mathbf{i}\alpha\rightarrow\mathbf{j}\beta\right)>0$.

\subsection{Bipolaron hopping}

Bipolaron hopping is a second order process\cite{tsigankov} where
a bipolaron $\langle\mathbf{i}\alpha,\mathbf{i}'\alpha'\rangle$ hops
to $\langle\mathbf{j}\beta,\mathbf{j}'\beta'\rangle$ via the virtual
dissociated states $(\mathbf{j}\beta,\mathbf{i}'\alpha)'$ or $(\mathbf{i}\alpha,\mathbf{j}'\beta')$.
Tsigankov and Efros\cite{tsigankov} argued that the hopping probability
is suppressed when the distance between the particles \emph{within}
a hopping pair is large, which reduces the biparticle conductivity
by orders of magnitude. This is not an issue in our case since the
size of our bipolarons is restricted to one lattice constant. 

The matrix element for the bipolaron hop is given by: \begin{equation}
A=\frac{<\mathbf{i}\alpha,\mathbf{i}'\alpha'|\hat{V}|\mathbf{j}\beta,\mathbf{i}'\alpha'><\mathbf{j}\beta,\mathbf{i}'\alpha'|\hat{V}|\mathbf{j}\beta,\mathbf{j}'\beta'>}{{E_{\mathbf{i}\alpha,\mathbf{i}'\alpha'}-E_{\mathbf{j}\beta,\mathbf{i}'\alpha'}}}\label{eq:tran-ampl}\end{equation}
 where the operator $\hat{V}$ is determined by Eq. (4.13) of Ref.\cite{alemott1995}
and has both diagonal and off-diagonal elements in phonon occupation
numbers. Each matrix element in (\ref{eq:tran-ampl}) contains a factor
$\exp{(-\alpha_{\mathrm{sp}}R)}$, which depends on the distance of
the hop. The dependence of $A$ on the hop length is therefore $\exp{[-\alpha_{\mathrm{sp}}(r_{\mathbf{ij}}+r_{\mathbf{i}'\mathbf{j}'})]}$. 

Another question which should be addressed is the frequency of the
hopping events which is determined by the pre-exponential factor of
the matrix elements. Here we are interested in the case when the single
polaron hopping is allowed so the binding energy of the pair, $\Delta$,
is small, $\Delta<<\omega,$ where $\omega$ is the phonon frequency.
We can therefore assume that $E_{\mathbf{i}\alpha,\mathbf{i}'\alpha'}-E_{\mathbf{j}\beta,\mathbf{i}'\alpha'}\simeq-\omega$
and the pre-exponential factor of $A$ is $t^{2}/\omega$, where $t$
is the frequency for the single polaron jump. Since we are interested
in the single phonon processes the second matrix element in (\ref{eq:tran-ampl})
must be diagonal in the phonon occupation numbers. The single polaron
jumps occur with the frequency $t\sim\omega$ so up to a numerical
factor of the order of 1 the pre-exponential factor of $A$ is $\sim\omega$,
similar as in the single polaron hop case. The bipolaron-hopping-rate
temperature independent pre-factor is therefore $\gamma_{0}^{\mathrm{bp}}e^{-2\alpha_{\mathrm{sp}}\left(r\mathbf{_{ij}}+r_{\mathbf{i}'\mathbf{j}'}\right)}$,
with $\gamma_{0}^{\mathrm{bp}}\approx\gamma_{0}$, and the  \emph{bipolaron
hopping rate} is given by: \begin{widetext} \begin{equation}
\gamma_{\langle\mathbf{i}\alpha,\mathbf{i}'\alpha'\rangle\langle\mathbf{j}\beta,\mathbf{j}'\beta'\rangle}^{\mathrm{bp}}=\gamma_{0}^{\mathrm{bp}}e^{-2\alpha_{\mathrm{sp}}\left(r\mathbf{_{ij}}+r_{\mathbf{i}'\mathbf{j}'}\right)}\begin{cases}
e^{-\frac{\Delta_{\langle\mathbf{i}\alpha,\mathbf{i}'\alpha'\rangle\langle\mathbf{j}\beta,\mathbf{j}'\beta'\rangle}}{T}} & \mathrm{;}\,\Delta_{\langle\mathbf{i}\alpha,\mathbf{i}'\alpha'\rangle\langle\mathbf{j}\beta,\mathbf{j}'\beta'\rangle}>0\\
1 & \mathrm{;}\,\Delta_{\langle\mathbf{i}\alpha,\mathbf{i}'\alpha'\rangle\langle\mathbf{j}\beta,\mathbf{j}'\beta'\rangle}\leq0\end{cases},\label{eq:bp-hop-prob}\end{equation}
\end{widetext} where $\langle\mathbf{i}\alpha,\mathbf{i}'\alpha'\rangle$
denotes the nearest neighbors and \begin{equation}
\begin{split}\Delta_{\langle\mathbf{i}\alpha,\mathbf{i}'\alpha'\rangle\langle\mathbf{j}\beta,\mathbf{j}'\beta'\rangle}= & \Delta H\left(\langle\mathbf{i}\alpha,\mathbf{i}'\alpha'\rangle\rightarrow\langle\mathbf{j}\beta,\mathbf{j}'\beta'\rangle\right)-\\
 & -E\left(x_{\mathbf{ij}}\textnormal{+}x_{\mathbf{i}'\mathbf{j}'}\right)\end{split}
,\end{equation}
is the many body energy difference between the final and initial states.

The total bipolaron instantaneous current for a particular system
configuration in the limit of vanishing electric field is similar
to the single polaron case,\begin{widetext} \begin{equation}
i^{\mathrm{bp}}=N_{\mathrm{eff}}\mu{}_{\mathrm{bp}}E\approx\gamma_{0}^{\mathrm{bp}}\frac{E}{T}\sum_{\langle\mathbf{i}\alpha,\mathbf{i}'\alpha'\rangle}\sum_{\langle\mathbf{j}\beta,\mathbf{j}'\beta'\rangle}^{\prime}\left(x_{\mathbf{ij}}+x_{\mathbf{i}'\mathbf{j}'}\right)^{2}e^{-2\alpha_{\mathrm{sp}}\left(r\mathbf{_{ij}}+r_{\mathbf{i}'\mathbf{j}'}\right)}e^{-\frac{\Delta H\left(\langle\mathbf{i}\alpha,\mathbf{i}'\alpha'\rangle\rightarrow\langle\mathbf{j}\beta,\mathbf{j}'\beta'\rangle\right)}{T}}.\label{eq:bp-current}\end{equation}
 \end{widetext}The first sum runs over all occupied bipolaron and
the second over all possible unoccupied bipolaron states with $\Delta H\left(\langle\mathbf{i}\alpha,\mathbf{i}'\alpha'\rangle\rightarrow\langle\mathbf{j}\beta,\mathbf{j}'\beta'\rangle\right)>0$.
For the sake of comparison with the single-polaron mobility Eq.(\ref{eq:bp-current})
also defines the effective dimensionless bipolaron mobility $\mu_{\mathrm{bp}}$
per single particle/hole.

\section{Results and discussion}

We calculate the polaron and bipolaron mobilities at any given $T$
by averaging equations (\ref{eq:sp-current}) and (\ref{eq:bp-current})
over Markov chains of the system configurations obtained by means
of MC simulations. Contrary to ref. {[}\onlinecite{tsigankov}{]}
a realistic MC dynamics, which includes correlated biparticle hops
in the external field, is \emph{not required}. Instead, unrestricted-length
single-particle hops were used to improve convergence. The details
of the used MC algorithm were described in ref.\cite{MerteljKabanov2007}. 

\begin{figure}[h!]
 \centering{}\includegraphics[scale=0.5]{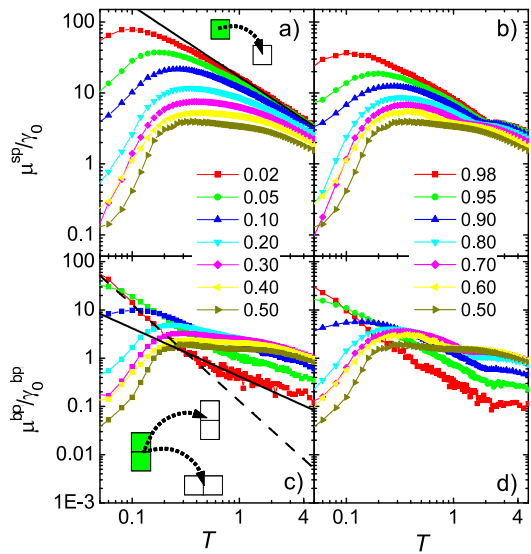} \caption{\label{fig:mobility} Temperature dependence of the single polaron
mobility, $\mu^{\mathrm{sp}}$, (a) and (b) and the effective bipolaron
mobility, $\mu^{\mathrm{bp}}$, (c) and (d) at $V_{\mathrm{JT}}=-1$
at different doping levels $n$. In the insets the corresponding hopping
processes are shown schematically. The solid and dashed black lines
show $1/T$ and $1/T^{2}$ slopes, respectively.}

\end{figure}

For the present calculations we used periodic boundary conditions
on a system size $L\times L=30\times30$ after carefully checking
for any size effects. The shape of the $V_{\mathrm{JT}}(\mathbf{i})$
was taken to be nonzero only for $\left|\mathbf{i}\right|=1$ and
was therefore specified by a single parameter $V_{\mathrm{JT}}(1,0)=V_{\mathrm{JT}}$.
To minimize further the computation time we limit the maximum hopping
distance in equations (\ref{eq:sp-current}) and (\ref{eq:bp-current}).
After a careful check for convergence we set $\alpha_{\mathrm{sp}}=0.5$
and selected the maximum hopping distance, $r_{\mathrm{C}}=8$. In
Fig. \ref{fig:mobility} we plot calculated temperature dependence
of mobilities at different dopings $n=N/L^{2}$, where $N$ is the
total number of polarons in the system and $n<0.5$ represents particle
and $n>0.5$ hole doping regions.

At high $T$ the second exponent in the sums (\ref{eq:sp-current})
and (\ref{eq:bp-current}) shows a negligible temperature dependence
so the dominant contribution to temperature dependence of both mobilities
comes from the $1/T$ prefactor. When the temperature is decreased
we first observe the suppression of mobilities due to the orbital
(pseudospin) ordering\cite{mertelj,MerteljKabanov2007} in the hole
doping region, $n\geq0.5$, which is most pronounced in the single
polaron mobility. The suppression is due to the high energy cost of
the hops with the pseudospin flip in the orbital ordered system which
makes half of the final states almost inaccessible.

\begin{figure*}[t]
 \centering{}\includegraphics[scale=0.8,angle=-90]{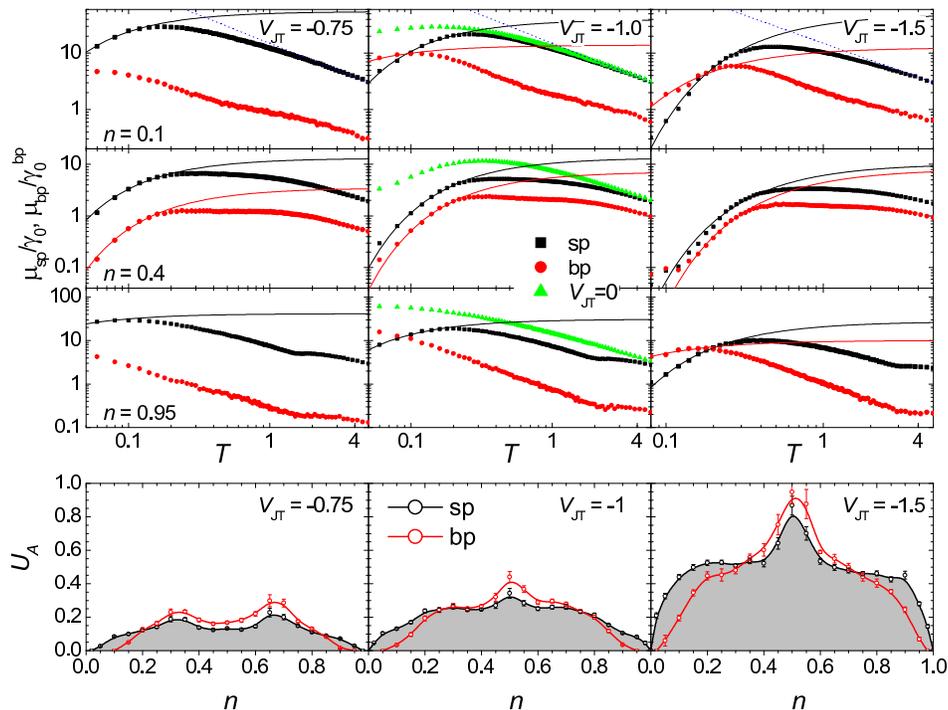}
\caption{\label{fig:Phase-diagrams}Polaron and bipolaron mobilities at three
doping levels: $n=0.1$, $0.4$, $0.95$ (top to bottom rows) and
three JT couplings: $V_{\mathrm{JT}}=-0.75$, $-1.0$, $-1.5$ (left
to right columns). For comparison the single-charge mobility in the
absence of the JT interaction is also shown. Thin full lines are low-$T$
Arrhenius fits and dotted lines show $T^{-1}$ slope. Panels at the
bottom show the activation energy, $U_{\mathrm{A}}$, as a function
of doping. }

\end{figure*}

With further decrease of temperature, the behavior of the mobility
at low particle and low hole doping becomes almost identical: The
single polaron mobility shows a maximum at any doping at the temperature
$T_{\mathrm{M}}^{\mathrm{sp}}$, which is doping dependent. In contrast,
the maximum is absent for the bipolaron mobility at low particle/hole
doping. The bipolaron mobility even starts to increase faster than
$1/T$ at low temperatures \emph{exceeding} the single polaron mobility
at low particle/hole doping and sufficiently strong attractive JT
interaction, $V_{\mathrm{JT}}\lesssim-1$ (see Fig. \ref{fig:Phase-diagrams}).
This is the most important result of our calculations. The explanation
of this effect follows from the the fact that all particles in the
cluster are paired, therefore formation of the clusters with odd number
of particles is highly energetically unfavorable. With increased doping
and strength of the attractive JT interaction also $\mu_{\mathrm{bp}}$
starts to show an inflection point at the temperature, $T_{\mathrm{M}}^{\mathrm{bp}}$,
which is always lower in the bipolaron case. Below $T_{\mathrm{M}}^{\mathrm{sp}}$
($T_{\mathrm{M}}^{\mathrm{bp}}$) both mobilities show an Arrhenius
temperature dependence (see Fig. \ref{fig:Phase-diagrams}). The activation
energies clearly show a \emph{smaller} energy barrier for bipolaron
hopping in comparison to single polaron hopping. 

In our simulation $\mu_{\mathrm{bp}}/\gamma_{0}^{\mathrm{bp}}$ exceeds
$\mu_{\mathrm{sp}}/\gamma_{0}$ at small doping and low temperature.
This however does not imply that the bipolaron mobility cannot dominate
the charge transport even at the intermediate dopings and for weaker
JT attractions. If the binding energy of a bipolaron is larger than
the phonon frequency, the single phonon processes which break the
bipolaron are forbidden. The prefactor $\gamma_{0}$ in the single
particle hopping probability (\ref{eq:hop-prob}) will be therefore
strongly suppressed\cite{AlexandrovKabanov1986} because more then
one phonon is necessary to conserve energy. On the other hand, the
bipolaron hopping processes do not require real bipolaron breaking.
As our simulations show (see bottom panels in Fig. \ref{fig:Phase-diagrams}),
the bipolaron hopping processes can have a lower energy barrier even
with a moderate JT attraction and can be assisted by a single phonon
leading to $\gamma_{0}^{bp}>\gamma_{0}$. This results in a further
suppression of the single polaron transport with respect to the bipolaron
transport.

In the hole doping region at low temperatures the system shows complete
orbital (pseudospin) order.\cite{MerteljKabanov2007} The effective
attractive JT interaction between holes governing the hopping process
then becomes isotropic. Since in this case the bipolaron mobility
also exceed the single-polaron one we can conclude that the anisotropy
of the attractive interaction is unimportant with respect to the domination
of the bipolaron transport at low $T$.

\section{Conclusions}

Using a new technique for the calculation of the transport properties
of polarons and bipolarons in the charge segregated state of doped
classical strongly correlated systems in the hopping regime, we analyzed
the electrical conductivity of a quasi-2D system with charge segregated
polarons in a textured state driven by the Coulomb interaction. The
problem of choosing the appropriate hopping probability, which has
been an issue in previous works\cite{tsigankov} has been sidestepped
by the fact that the size of our bipolaron is small (one lattice constant).
We have shown that at high temperature the mobility of carriers increases
with decreasing temperature as $T^{-1}$. Single-polaron mobility
shows a maximum at the characteristic temperature where clusters start
to form. Below this temperature, the single-polaron conductivity is
exponentially suppressed because the single-polaron hopping requires
dissociation of the polaron from a cluster. On the other hand, the
bipolaron transport shows a weaker suppression, which starts at a
lower temperature, and thus becomes the dominant transport channel
at low temperatures. Apart from resistivity, the Peltier effect, the
Nernst effect, the Hall effect and particularly the Lorentz number
are all expected to exhibit a low-temperature crossover when bipolaron
transport becomes dominant, so the model predictions are eminently
verifiable by experiment.
\begin{acknowledgments}
This work has been supported by Slovenian Research Agency and CONACYT
Mexico.\end{acknowledgments}

\end{document}